\documentclass[aps,prl,twocolumn,groupedaddress]{revtex4-1}
\usepackage{amsmath}
\usepackage{graphicx}
\begin{document}

\preprint{CP3-Origins-2017-040}
\preprint{MITP/17-063}

\title{Elastic $I=3/2$ $p$-wave nucleon-pion scattering amplitude 
and the $\Delta(1232)$ resonance from $N_{\mathrm{f}}=2+1$ lattice QCD}


\author{Christian~Walther~Andersen}
\author{John~Bulava}
\affiliation{Dept. of Mathematics and Computer Science and CP3-Origins, University of Southern Denmark,
Campusvej 55, 5230 Odense M, Denmark}
\author{Ben~H\"{o}rz}
\affiliation{PRISMA Cluster of Excellence and Institut f\"{u}r Kernphysik,
University of Mainz, Johann-Joachim-Becher-Weg 45, 55099 Mainz, Germany}
\author{Colin~Morningstar}
\affiliation{Department~of~Physics, Carnegie~Mellon~University, 
Pittsburgh, PA~15213, USA}
%\thanks{}


\date{\today}

\begin{abstract}
We present the first direct determination of 
	meson-baryon resonance parameters from a scattering amplitude calculated 
	using lattice QCD. In 
	particular, we calculate the elastic $I=3/2$, 
	$p$-wave nucleon-pion amplitude on a single ensemble of $N_{\mathrm{f}}=2+1$ 
	Wilson-clover fermions with $m_{\pi}=280\mathrm{MeV}$ and 
	$m_{K}=460\mathrm{MeV}$. At these  
	quark masses, the $\Delta(1232)$ resonance pole is found close to the $N-\pi$ 
	threshold and a Breit-Wigner fit to the amplitude gives 
	 $g^{\mathrm{BW}}_{\Delta N\pi}=19.0(4.7)$ in agreement with phenomenological 
	 determinations. 
\end{abstract}

%\keywords{}

\maketitle

\section{Introduction\label{s:intro}}
Accurate and precise predictions of hadron-hadron scattering amplitudes 
from first principles are desirable for many phenomenological applications. 
While lattice QCD has been successful in calculating many single-hadron 
properties, hadron-hadron scattering amplitudes have been more difficult.  
Since lattice QCD simulations are performed in Euclidean time and finite 
volume, real-time infinite-volume scattering amplitudes cannot be calculated 
directly~\cite{Maiani:1990ca}. Instead, the finite volume may be 
exploited to determine scattering amplitudes using the 
shift of interacting two-hadron energies from their non-interacting 
values~\cite{Luscher:1990ux}. 

However, these calculations have been hampered by the difficulty in evaluating  correlation functions with two-hadron interpolating operators.  
Thanks to algorithmic advances in the treatment of all-to-all 
quark propagators~\cite{Peardon:2009gh,Morningstar:2011ka} and increasing computational resources, 
lattice QCD studies of scattering amplitudes have undergone substantial recent 
progress.
As reviewed in (e.g.) Ref.~\cite{Briceno:2017max}, many calculations of 
resonant meson-meson amplitudes have been performed. Elastic meson-meson 
scattering amplitudes are therefore quickly entering an era of 
high precision, 
while first progress has been made on amplitudes with multiple coupled 
meson-meson scattering channels and on amplitudes coupled to external currents~\cite{Alexandrou:2017mpi,Bali:2015gji,Fu:2016itp,Feng:2014gba,Feng:2010es,
Orginos:2015aya,Beane:2011sc,Pelissier:2012pi,Aoki:2011yj,
Moir:2016srx,Briceno:2016mjc,Briceno:2016kkp,Dudek:2016cru,Wilson:2015dqa,
Wilson:2014cna,Dudek:2012xn,Dudek:2012gj,Lang:2016jpk,Lang:2015hza,
Lang:2014yfa,Mohler:2013rwa,Prelovsek:2013ela,Mohler:2012na,Lang:2012sv,
Lang:2011mn,Guo:2016zos,Helmes:2017smr,Liu:2016cba,Helmes:2015gla,
Bulava:2016mks}. 

The situation with meson-baryon scattering amplitudes is considerably less 
advanced. There are calculations of non-resonant amplitudes and scattering 
lengths~\cite{Fukugita:1994ve, Meng:2003gm, Torok:2009dg, 
Detmold:2015qwf}, as well as first steps towards resonant $N-\pi$ 
amplitudes~\cite{Lang:2012db,Lang:2016hnn}. Nonetheless, 
to date published determinations of meson-baryon 
resonance parameters from amplitudes calculated using lattice QCD are lacking.  
Unpublished preliminary progress toward a calculation of the 
amplitude considered in this work was communicated privately in Fig.~17 of 
Ref.~\cite{Mohler:2012nh}.

The $\Delta(1232)$ is the lowest-lying baryon resonance, but remains 
phenomenologically interesting. For instance, as discussed in Ref.~\cite{Alvarez-Ruso:2017oui} 
 nucleon-$\Delta(1232)$ transition form factors are an important 
 phenomenological 
input for neutrino-nucleus scattering experiments such as NO$\nu$A and DUNE.
The scattering amplitude calculated in this work is 
a required first step in the calculation of such form factors using lattice QCD. 

There are many difficulties associated with meson-baryon scattering 
amplitudes. First, the exponential degradation in the signal-to-noise ratio 
is typically 
worse in correlation functions containing baryon interpolators than in the 
pure-meson sector. Second, the additional valence quark results in increased 
computational and storage costs, as well as a proliferation of the necessary 
Wick contractions. Furthermore, resonant meson-baryon 
amplitudes require `annihilation diagrams' which are present in 
correlation functions between single-baryon and meson-baryon interpolators. 
Finally, nonzero baryon spin complicates the construction 
of irreducible meson-baryon operators~\cite{Morningstar:2013bda} and the 
extraction of scattering amplitudes~\cite{Morningstar:2017spu}.

Despite these difficulties, we present here the first lattice QCD calculation of the resonant $I=3/2$, 
$N-\pi$ amplitude in the elastic region on a single ensemble of gauge field 
configurations with  
$N_{\mathrm{f}}=2+1$ dynamical flavors of Wilson-clover fermions 
generated as part of the Coordinated Lattice Simulations (CLS) 
consortium~\cite{Bruno:2014jqa}. Although this ensemble is at unphysically 
heavy (degenerate) light quark mass corresponding to 
$m_{\pi}=280\mathrm{MeV}$, we observe an analogue of the $\Delta(1232)$ 
resonance close to $N-\pi$ threshold.   

Using a variety of moving frames~\cite{Rummukainen:1995vs,Gockeler:2012yj}, we employ  
finite-volume energies to determine the amplitude at six points 
in the elastic region. The energy dependence of the resultant amplitude is 
well-described by a Breit-Wigner resonance shape, yielding 
$m_{\Delta} = 1344(20)\mathrm{MeV}$ and 
$g^{\mathrm{BW}}_{\Delta N\pi}=19.0(4.7)$ 
already at our heavier-than-physical light quark masses.

This letter is organized as follows.  We first detail the lattice
QCD ensemble employed here, our method for extracting the spectrum, and the 
subsequent determination of the amplitude. This is followed by the presentation and analysis of  
the results. Finally, we close with conclusions and an outlook.

\section{\label{s:meth}Lattice QCD methods}

\emph{Ensemble details}: The gauge field ensemble employed here is from the CLS 
consortium~\cite{Bruno:2014jqa,Bali:2016umi}, which has generated a large set 
of $N_{\mathrm{f}}=2+1$ flavor ensembles at several lattice spacings and quark 
masses.  
\begin{table}
	\centering
	\begin{tabular}{c|c|c|c|c|c|c}
		\hline
		ID & $\beta$ & $a(\mathrm{fm})$ & $L^3 \times T$  & $m_{\pi},\, m_{K} (\mathrm{MeV})$ & 
		$N_{\mathrm{conf}}$ & $N_{t_0}$   \\ 
		\hline
		N401 & $3.46$ & $0.0765$ & $48^3\times128$ & $280,\, 460$ & $275$ & $2$ 
	\end{tabular}
	\caption{\label{t:ens}Parameters of the CLS ensemble used in this work. After the ensemble ID in the first column, we list 
	the gauge coupling, lattice spacing and dimensions, pseudoscalar meson masses,  number of gauge configurations, and number of source times.
	 A precise lattice spacing determination can be 
	found in Ref.~\cite{Bruno:2016plf}.
	} 
\end{table}
This single ensemble is detailed in Tab.~\ref{t:ens} and does not have the quark masses set to their physical values
but belongs to a quark mass trajectory where $2m_{\ell} + m_{s}$ is kept fixed as 
$m_{\ell}=m_{u}=m_{d}$ is lowered to its physical value. Therefore we have both 
$m_{\pi} > m_{\pi}^{\mathrm{phys}}$ and $m_{K} < m_{K}^{\mathrm{phys}}$.  

This ensemble also employs the open temporal boundary conditions of 
Ref.~\cite{Luscher:2011kk}. In order to ensure a Hermitian matrix of correlation functions, 
our interpolating operators are always separated from the temporal boundaries 
by at least $t_{\mathrm{bnd}}$, where $m_{\pi}t_{\mathrm{bnd}} = 3.5$. Using 
the zero-momentum single-pion correlation function, which is the most precisely 
determined 
correlation function, we have demonstrated that this separation is sufficient 
to reduce temporal boundary effects below the statistical precision. Temporal 
boundaries are therefore neglected in all subsequent analysis. 

\emph{Correlation functions}: In order to efficiently treat the all-to-all quark propagators required in two-hadron correlation functions, we employ the 
stochastic LapH method~\cite{Morningstar:2011ka}. While brute-force 
calculation of the entire quark propagator is intractable, this method 
projects it onto 
a low-dimensional subspace spanned by $N_{\mathrm{ev}}$ eigenmodes of the 
stout link-smeared~\cite{Morningstar:2003gk} 
gauge-covariant 3-D Laplace operator. This projection is a form of quark smearing, a common technique used to reduce unwanted excited state contamination in temporal correlation functions. 

The stochastic LapH method~\cite{Morningstar:2011ka} then introduces stochastic estimators for the smeared-smeared quark propagator $\mathcal{Q}(x,y)$ in this 
subspace spanned by time (`T'), spin (`S'), and Laplacian eigenvector (`L') 
indices. The variance 
of these estimators may be improved via dilution~\cite{Foley:2005ac}. In each 
index we shall either consider full dilution, denoted `$\mathrm{F}$', or $n$ 
uniformly interlaced dilution projectors, denoted `$\mathrm{I}n$'. 
Furthermore, it is beneficial to employ different dilution schemes for `fixed' 
quark lines, where $x_0 \ne y_0$, and for `relative' quark lines, where $x_0 = y_0$. Fixed and relative dilution schemes are denoted by the subscripts `F' and `R', respectively. 
\begin{table}
	\centering 
	\begin{tabular}{c|c|c|c|c}
\hline 
		$(\rho, n_{\rho})$  &$N_{\mathrm{ev}}$ & dilution scheme & $N^{\mathrm{fix}}_{R}$ & $N^{\mathrm{rel}}_{R}$ \\
\hline 
		 $(0.1, 25)$ & $320$ & $(\mathrm{TF,SF,LI16})_F\, (\mathrm{TI8, SF, 
		 LI16})_R$ & $5$ & $1$    \\ 
	\end{tabular}
	\caption{\label{t:laph}Parameters of the stochastic LapH implementation used 
	in this work.  $(\rho,n_{\rho})$ are the stout link smearing 
	parameters, $N_{\mathrm{ev}}$ the number of Laplacian eigenvectors, and 
	$N_R$ the number of 
	independent stochastic sources 
	quark lines for fixed and relative quark lines. Notation for the dilution scheme is explained in the text.}
\end{table}
Information on the stochastic LapH implementation is given in 
Tab.~\ref{t:laph}. This scheme together with the $N_{t_0}=2$ source times 
for our fixed quark lines results in $N_D = 1152$ Dirac matrix inversions per 
configuration. Using this algorithm, all required Wick contractions are 
evaluated as described in Ref.~\cite{Lang:2012db} while only a single 
permutation of the stochastic quark line estimates is employed. To increase 
statistics
we average over all irrep rows and a subset of equivalent total momenta $\textbf{P}$. 

\emph{Energy calculation}: Shifts of the finite-volume $N-\pi$ energies 
from their noninteracting values are calculated directly by fitting the 
ratios~\cite{Bulava:2016mks}
\begin{align}\label{e:ratio}
	R_n(t) &= \frac{\hat{C}_{n}(t)}{C_{\pi}(\textbf{p}_{\pi,n}^2, t) \, 
	C_N(\textbf{p}_{N,n}^{2},t)}, 
\\\nonumber	
	\hat{C}_n &= (v_n(t_0,t_d), C(t) v_n(t_0, t_d)) 
\end{align}
to the ansatz $R_n(t) = A\mathrm{e}^{-\Delta E_n t }$. In Eq.~\ref{e:ratio}
$C(t)$ is a correlator matrix in a particular irreducible representation 
(irrep) and $v_n(t_0, t_d)$ an eigenvector from the generalized eigenvalue 
problem (GEVP) $C(t_d) v_n = \lambda_n C(t_0) v_n$. $C_{\pi}$ and $C_N$ 
are single-pion and single-nucleon correlation functions (respectively) with momenta equal to 
those in the noninteracting $N-\pi$ level closest to $E_n$. While the 
GEVP enables the extraction of excited states, it is also practically 
advantageous to enhance ground-state overlap in correlators with 
significant mixing between the operators and eigenstates. 

We include one single-site (smeared) $\Delta$ interpolating operator and several 
nucleon-pion interpolators in the GEVP for each irrep resulting in correlation 
matrices of dimension $N_{\mathrm{op}} \lesssim 5$.   
We employ the ground state in each irrep in our analysis, as well as a single 
precisely determined excited state. With our current level of statistics, other levels in the elastic region have insufficient statistical precision to 
constrain the amplitude. 

\begin{figure}
	\includegraphics[width=0.5\textwidth]{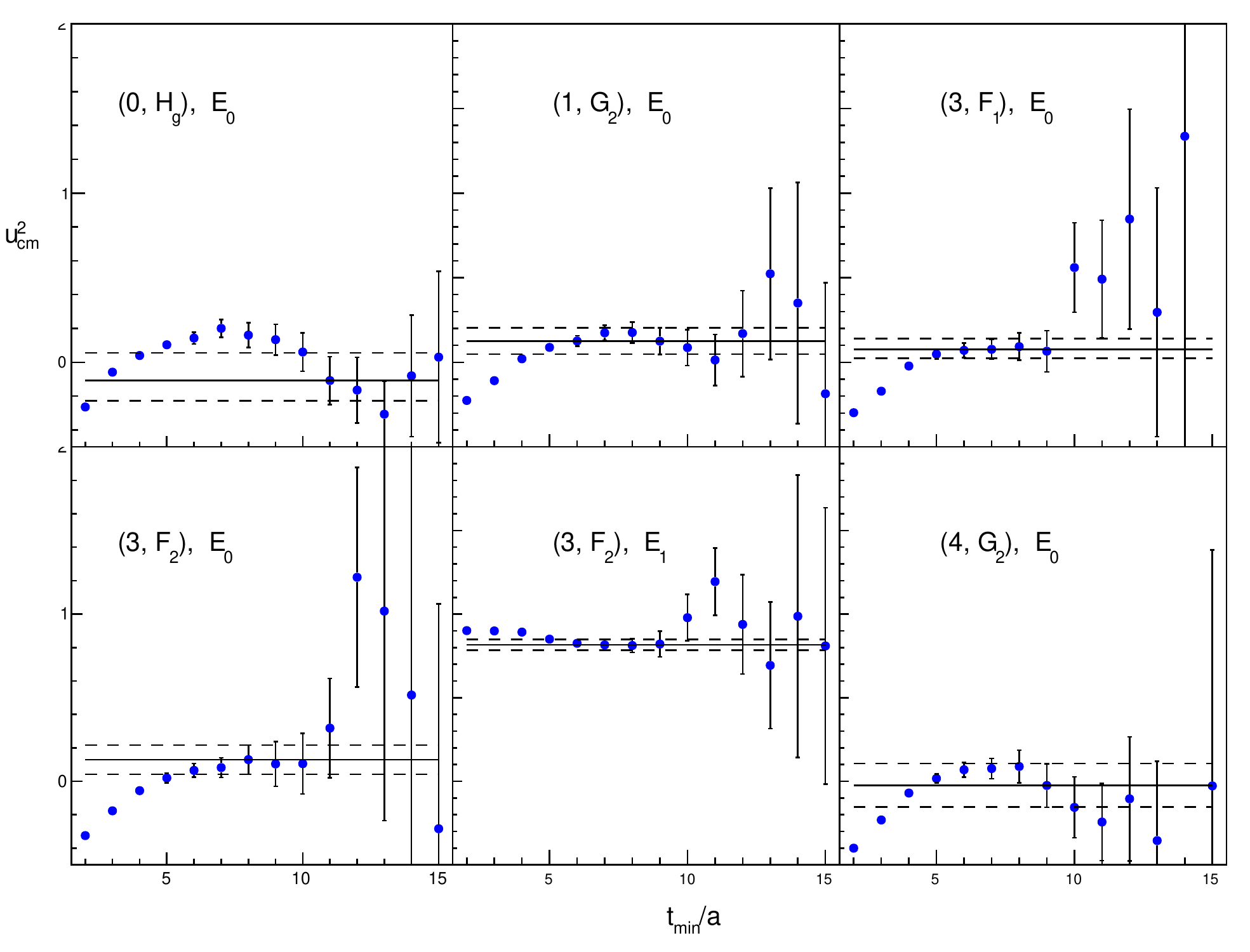}
	\caption{\label{f:tmin} Variation of $t_{\mathrm{min}}$, the lower end of 
	the range for single-exponential fits to the correlator ratio in 
	Eq.~\ref{e:ratio}. The vertical axis is the dimensionless center-of-mass momentum $u_{\mathrm{cm}}^2 = L^2q^2_{\mathrm{cm}}/(2\pi)^2$. Shown here are the ground state energies from each of the five irreps $(\textbf{P}^2,\Lambda) = \{ 
	(0, H_g),\, (1, G_2),\, (3, F_1),\, (3,F_2),\,(4, G_2)\}$ and a single first excited state from the $(3,F_2)$ irrep. For each fit $t_{\mathrm{max}}=25a$ while the dashed lines indicate the chosen fit.} 
\end{figure}

Effects due to variation of $(t_0,t_d)$ and $N_{\mathrm{op}}$
are not visible with our current statistical precision.
Furthermore, as seen in Fig.~\ref{f:tmin}, we 
choose fit ranges $[t_{\mathrm{min}}, t_{\mathrm{max}}]$ with $t_{\min}$ large 
enough so that the systematic error due to unwanted excited state contamination is smaller than the 
statistical error. It should be noted that the excited state contamination 
in $R(t)$ may be non-monotonically decreasing, leading to `bumps' in the $t_{\min}$-plots shown in Fig.~\ref{f:tmin}. Nonetheless, the overall magnitude of 
the excited state contamination is considerably smaller than in 
single-exponential fits to just the numerator of Eq.~\ref{e:ratio}.
Furthermore, the chosen $t_{\mathrm{min}}$ values lie in the plateau region for 
individual effective masses of both numerator and denominator, so we are 
confident that $R_n(t)$ behaves asymptotically for $t\ge t_{\mathrm{min}}$. 

Although multihadron correlation functions containing baryons are more 
computationally intensive than those with just mesons, the overall measurement cost is still dominated by the Dirac matrix inversions, which we perform efficiently using the \texttt{DFL\_SAP\_GCR} solver in \texttt{openQCD}~\cite{oqcd}.
However, the baryon functions defined in Eq.~23 of Ref.~\cite{Morningstar:2011ka} are the dominant storage cost. 

\emph{Amplitude calculation}: A variant of the methods of 
Refs.~\cite{Luscher:1990ux, Rummukainen:1995vs} detailed in 
Refs.~\cite{Gockeler:2012yj,Morningstar:2017spu} is applied to relate 
finite-volume 
$N-\pi$ energies to the infinite-volume elastic scattering amplitude. For each total momentum $\textbf{P}$ and irrep $\Lambda$, these 
relations are given as determinant conditions of the form
\begin{align}\label{e:det}	
	\mathrm{det}(\hat{K}^{-1}-B^{(\textbf{P},\Lambda)}) = 0, 
\end{align}
which hold up to exponentially suppressed residual 
finite volume effects. The determinant is taken over indices corresponding 
to the total angular momentum $J$, total orbital angular momentum $\ell$, total spin $S$,  and an 
occurrence index $n$ labelling multiple occurrences of the partial wave in the
irrep. The (infinite dimensional) matrix $B$ depends on the irrep and encodes the reduced 
symmetries of the finite volume. It is diagonal in $S$ but (in general) 
dense in all other indices. 
Expressions for all required elements of $B$ up to $J=13/2$ and $\ell=6$ are 
given in Ref.~\cite{Morningstar:2017spu}. $\hat{K}$ is diagonal in $J$, 
equal to the identity in $n$, and 
is related to the usual $K$-matrix via 
$K_{\ell S;\ell'S'}^{-1} = q_{\mathrm{cm}}^{-(\ell+\ell'+1)} 
\hat{K}_{\ell S;\ell'S'}^{-1}$ where $q_{\mathrm{cm}}$ is the center-of-mass 
momentum. 

For this first calculation we only include irreps in which the $J^{\eta}=3/2^{+}$, 
$p$-wave is the lowest contributing partial wave~\cite{Gockeler:2012yj}, namely 
 $(\textbf{P}^2,\Lambda) = \{ 
(0, H_g),\, (1, G_2),\, (3, F_1),\, (3,F_2),\,(4, G_2)\}$.
In addition to ignoring  
the exponential finite volume effects in Eq.~\ref{e:det},  contributions from 
higher 
$\ell>1$ are expected to be negligible based on 
threshold angular-momentum suppression. We assess the effect of this 
truncation to $\ell =1$ by performing 
a fit also including all $\ell=2$ contributions, namely the $J^{\eta} = 3/2^-$ and $5/2^{-}$ partial waves. For this fit with the $\ell=2$ waves, we additionally include the ground state energy in the $(0,H_u)$ irrep where the $3/2^{+}$ wave does not contribute, but both the $3/2^{-}$ and $5/2^{-}$ are present.

\section{Results\label{s:res}} 

\begin{figure}
	\includegraphics[width=0.5\textwidth]{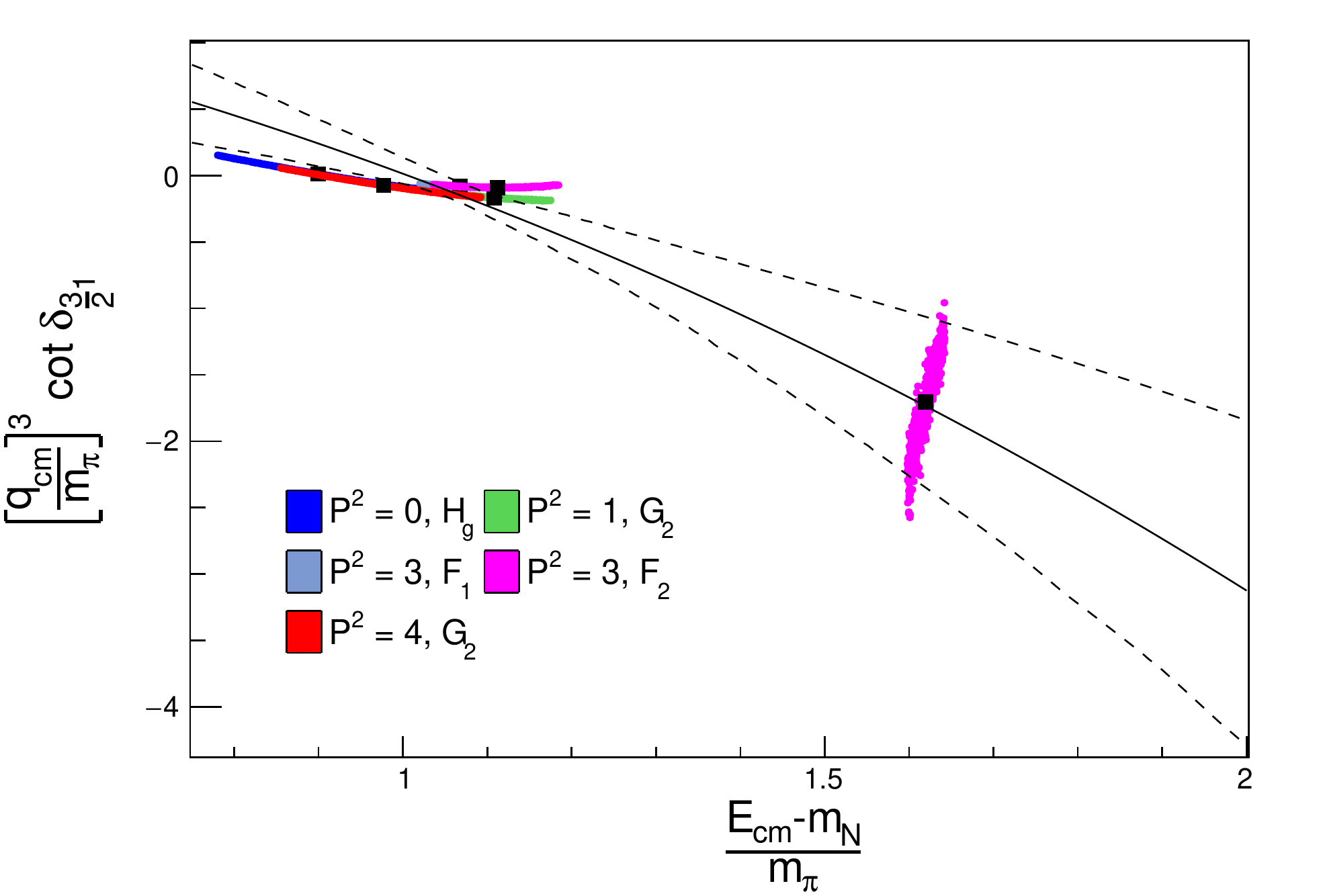}
	
	\includegraphics[width=0.5\textwidth]{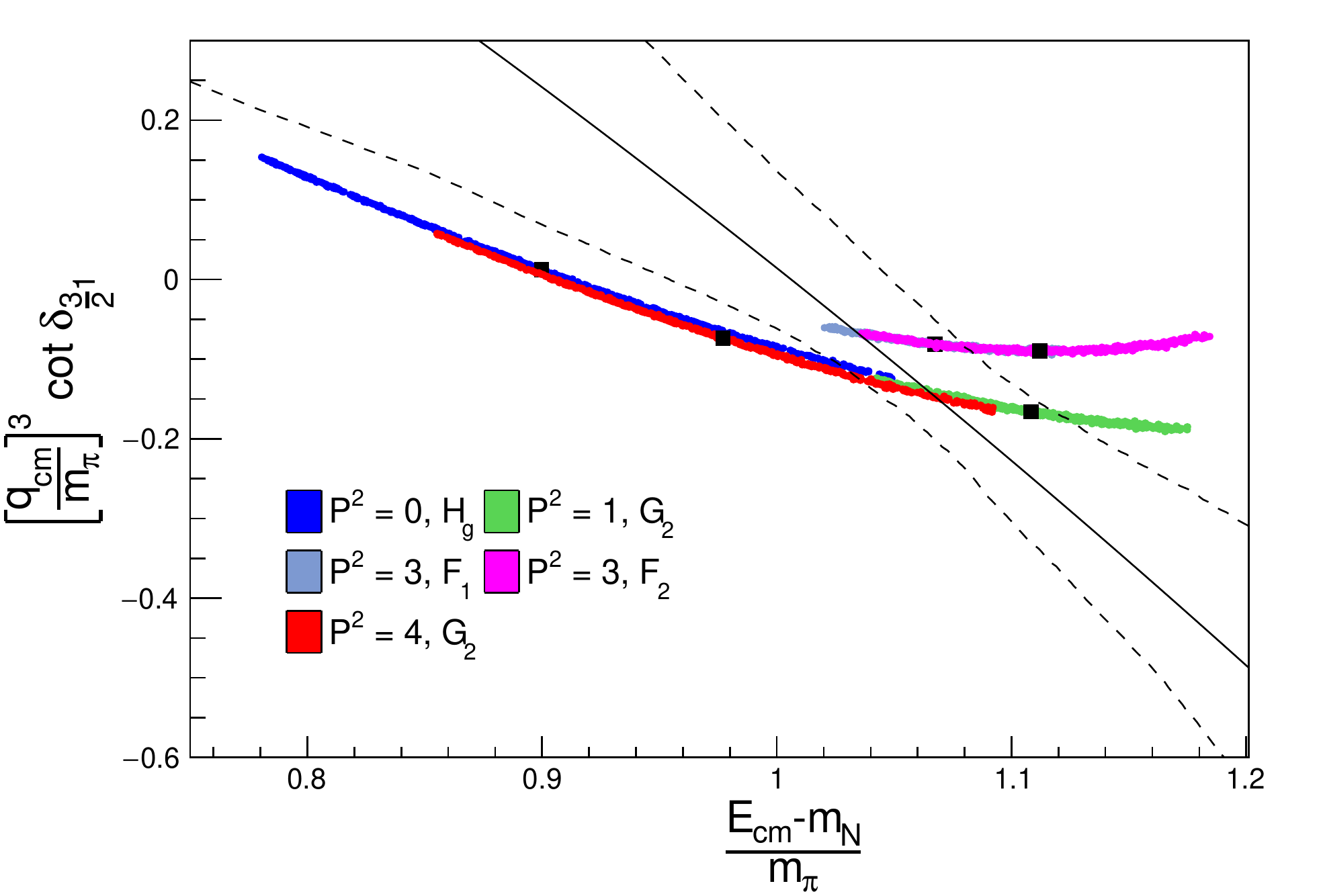}
	\caption{\label{f:p3cot}The real part of the inverse scattering amplitude 
	for $I=3/2$, $p$-wave elastic $N-\pi$ scattering. Different colors indicate different energy levels, for each of which a colored circle is plotted for each of the central 68\% of bootstrap samples and a black square indicates the mean 
	value. The solid and dotted lines denote the mean values and bootstrap errors,
	respectively, 
	for the Breit-Wigner fit described in the 
	text. The lower panel focusses on the resonance region shown in the upper 
	panel.} 
\end{figure}

Results for the $I=3/2$, $p$-wave elastic $N-\pi$ scattering amplitude are 
presented in Fig.~\ref{f:p3cot}, where the (rescaled) real part of the inverse 
amplitude $(q_{\mathrm{cm}}/m_{\pi})^3 \, \cot \delta_{\frac{3}{2}1}$ is shown 
as a function of the center-of-mass energy $E_{\mathrm{cm}}$. This quantity is 
smooth near the elastic $N-\pi$ threshold and, unlike the scattering phase 
shift, can describe both near-threshold resonances and bound states.
However it is a highly nonlinear function of  $E_{\mathrm{cm}}$, so that conventional horizontal and vertical error bars would be significantly correlated.
Instead of these, in Fig.~\ref{f:p3cot} for each energy we display one point 
for each of the central $68\%$ of bootstrap samples. This therefore gives a 
visual representation of the 1-$\sigma$ confidence interval for each point in 
this two-dimensional plot.
Finally, in Fig.~\ref{f:p3cot} $\tilde{E}_{\mathrm{cm}} = (E_{\mathrm{cm}}-
m_{N})/m_{\pi}$ is shown on the horizontal axis so that the elastic region is given by $1 < \tilde{E}_{\mathrm{cm}} < 2$. 

We describe the energy dependence of this amplitude with a Breit-Wigner shape 
\begin{align}\label{e:bw}
	\left(\frac{q_{\mathrm{cm}}}{m_{\pi}}\right)^{3}\cot \delta_{\frac{3}{2}1} = 
	\left(\frac{m^2_{\Delta}}{m^2_{\pi}} - \frac{E^2_{\mathrm{cm}}}{m_{\pi}^2}\right)
	\frac{6\pi E_{\mathrm{cm}}}{(g^{\mathrm{BW}}_{\Delta N \pi})^2 m_{\pi}}
\end{align}
with fit parameters $m_{\Delta}/m_{\pi}$ and $g^{\mathrm{BW}}_{\Delta N \pi}$. 
The fit is 
performed using the method of~\cite{Morningstar:2017spu} in which the residuals in the correlated-$\chi^2$ are taken to be 
\begin{align*}
	\Omega(\mu,A) = \frac{\mathrm{det}(A)}{\mathrm{det}\left[(\mu^2 + AA^{\dagger})^{1/2}\right]}, 
\end{align*}
where $A =  \hat{K}^{-1}-B^{(\textbf{P},\Lambda)}$ from Eq.~\ref{e:det}. We 
take $\mu =1$, although fit parameters do not vary outside their 
statistical errors when going from $\mu=1$ to $\mu=10$.

The results of this fit (which neglects $\ell > 1$ partial waves) are 
\begin{align*}
	\frac{m_{\Delta}}{m_{\pi}} = 4.738(47), \quad g^{\mathrm{BW}}_{\Delta N \pi} 
	= 19.0(4.7),
	\quad \chi^2/\mathrm{d.o.f.} = 1.11,
\end{align*}
where the errors are statistical only. While our small number of data points 
makes fits to other parametrizations difficult, we can attempt to describe this partial wave in a nonresonant manner by truncating the effective range 
expansion at leading order, yielding the one-parameter fit form 
\begin{align*}
	\left(\frac{q_{\mathrm{cm}}}{m_{\pi}}\right)^{3}\cot \delta_{\frac{3}{2}1}
	 &= \frac{1}{m_{\pi}^3 a_{{\frac{3}{2}1}}^3}.  
\end{align*}
This fit gives $(m_{\pi} a_{{\frac{3}{2}1}})^{-3} = -0.099(14)$ with $\chi^2/\mathrm{d.o.f.} = 2.50$, indicating a poorer description of the data compared to 
the Breit-Wigner form of Eq.~\ref{e:bw}. 

%With more data points and increased 
%statistics, the effect of different parametrizations could be studied in more
%detail. In particular, employing the first two terms in the effective range
%expansion would enable an alternative determination of the pole position 
%according to the method of e.g. Ref.~\cite{Lang:2015hza}. Unfortunately, with our current data such fits
%are insufficiently constrained. 

We can also assess the impact of the 
 $3/2^{-}$ and $5/2^{-}$ $d$-waves which are present in 
the irreps with nonzero total momenta. In addition to the six energies included in the previous fits, we add the ground state in the total zero momentum $H_u$ channel, where these two waves are the lowest contributing partial waves. 
Although we only have seven energy 
levels, we nonetheless perform a four-parameter fit including the leading term in the effective range 
expansion for each of these additional waves 
\begin{align*}
	\left(\frac{q_{\mathrm{cm}}}{m_{\pi}}\right)^{5}\cot \delta_{\frac{3}{2}2}
	 = \frac{1}{m_{\pi}^5 a_{{\frac{3}{2}2}}^5}, \quad 
	\left(\frac{q_{\mathrm{cm}}}{m_{\pi}}\right)^{5}\cot \delta_{\frac{5}{2}2}
	 = \frac{1}{m_{\pi}^5 a_{{\frac{5}{2}2}}^5}  
\end{align*}
together with the parametrization of Eq.~\ref{e:bw} for the $3/2^{+}$ $p$-wave. 
The results of this fit are 
\begin{align*}
	&\frac{m_{\Delta}}{m_{\pi}} = 4.734(56), \qquad g^{\mathrm{BW}}_{\Delta N \pi} = 19.0(7.4),
\\
	&(m_{\pi} a_{{\frac{3}{2}2}})^{-5} = 0.00(10), \quad 
	(m_{\pi} a_{{\frac{5}{2}2}})^{-5} = 0.00(12), \\ 
	&\chi^2/\mathrm{d.o.f.} = 4.17.
\end{align*}
The values for $m_{\Delta}$ and 
$g^{\mathrm{BW}}_{\Delta N \pi}$ are consistent with those obtained from truncating to $\ell =1$,  
confirming our insensitivity to these $\ell=2$ waves.

Since $m_N/m_{\pi} = 3.732(56)$, there is no significant difference between $m_{\Delta}$ and the elastic threshold at 
$E_{\mathrm{th}}/m_{\pi} = 1 + m_{N}/m_{\pi}$. By employing the scale determination of 
Ref.~\cite{Bruno:2016plf} we obtain a mass in physical units of 
$m_{\Delta} = 1344(20)\mathrm{MeV}$, where the error on the scale has been 
combined in quadrature. 
It is worth 
emphasising that the quark masses for this ensemble are tuned to satisfy 
$\mathrm{Tr} M = 2m_{l} + m_{s} = (\mathrm{Tr} M)^{\mathrm{phys}}$ as $m_l$ is lowered to its physical value, in contrast with the more standard choice where 
$m_s = m_s^{\mathrm{phys}}$ for all values of $m_l$.  

Comparison of $g^{\mathrm{BW}}_{\Delta N \pi}$ can be made to experiment 
using the experimental values $m^{\mathrm{exp}}_{\Delta}\approx 1232\mathrm{MeV}$ and 
$\Gamma^{\mathrm{exp}} \approx 117\mathrm{MeV}$ from Ref.~\cite{Patrignani:2016xqp} and the relation 
$\Gamma^{\mathrm{BW}} =  \frac{(g^{\mathrm{BW}}_{\Delta N \pi})^2 q_{\Delta}^{3}}{6\pi m_{\Delta}^2}$, where $q_{\Delta}$ is the center-of-mass momentum 
corresponding to the resonance mass. Such a comparison yields  $g^{\mathrm{BW},\mathrm{exp}}_{\Delta N \pi} \approx 16.9$, in agreement with our result.  

An alternative convention for the $\Delta N\pi$-coupling is provided by 
leading-order chiral effective theory~\cite{Pascalutsa:2005vq}, which defines 
$\Gamma$ as 
\begin{align*}
	\Gamma = \frac{(g^{\mathrm{LO}}_{\pi N\Delta})^2}{48\pi m_{N}^2}\frac{E_N + m_N}{E_N + E_{\pi}}q_{\Delta}^3,
\end{align*}
where $E_N$ and $E_\pi$ are the energies of the nucleon and pion, respectively, with momenta equal to $q_{\Delta}$. Using our calculated values for the resonance parameters and $m_N/m_{\pi}$ gives 
\begin{align*}
	g^{\mathrm{LO}}_{\pi N\Delta} = 37.1(9.2).
\end{align*}
Our result for $g^{\mathrm{LO}}_{\Delta N\pi}$ can be compared to previous lattice estimates 
using Fermi's Golden Rule from 
Refs.~\cite{Alexandrou:2015hxa, Alexandrou:2013ata},
which give  $g^{\mathrm{LO}}_{\Delta N\pi}=23.7(0.7)(1.1)$ at $m_{\pi} = 180\mathrm{MeV}$ and 
$g^{\mathrm{LO}}_{\Delta N\pi}=26.7(0.6)(1.4)$ at $m_{\pi} = 350\mathrm{MeV}$. We can also 
compare to a phenomenological extraction employing LO nucleon-pion effective
field theory~\cite{Pascalutsa:2005vq} yielding $g^{\mathrm{LO}}_{\Delta N\pi}=29.4(4)$, and 
 a phenomenological $K$-matrix analysis~\cite{Hemmert:1994ky} yielding $g^{\mathrm{LO}}_{\Delta N\pi}= 28.6(3)$.

\section{Conclusions and outlook\label{s:conc}}

 This work presents the first lattice determination of meson-baryon 
 resonance parameters directly from the scattering amplitude. It builds on 
 demonstrably successful 
 algorithms from the meson-meson sector~\cite{Bulava:2016mks}, in particular 
 the stochastic LapH method~\cite{Morningstar:2011ka}, which reduces 
 computational and storage costs for multihadron correlation functions 
 containing baryons significantly compared to the distillation 
 approach of Ref.~\cite{Peardon:2009gh}. 

 The $I=3/2$, $p$-wave, elastic $N-\pi$ scattering amplitude is calculated here 
 on a single ensemble of gauge configurations, and thus the usual lattice 
 systematic errors due to finite lattice spacing and 
 unphysical quark masses are not addressed. Furthermore while the magnitude of 
 the exponentially suppressed finite volume effects indicated in 
 Eq.~\ref{e:det} is presumably insignificant, this has not been checked 
 explicitly. 

 This first analysis also avoids the influence of $\ell=0$ partial wave
 mixing in Eq.~\ref{e:det} by judiciously choosing irreps where this 
 wave is absent and the $J=3/2$ $p$-wave is the leading contribution. Future 
 work will include also irreps where the corresponding $s$-wave is present, 
 which can be analyzed as described in Ref.~\cite{Morningstar:2017spu}. These additional 
 finite volume energies 
 will better constrain the energy dependence of the amplitude and enable a 
 more precise  
 analysis of higher partial wave contributions. Furthermore, this calculation
 employs only a single permutation of stochastic quark line estimates. 
 Additional `noise orderings' may significantly improve the statistical 
 precision.

 Furthermore, these results will soon be complemented by measurements on other
 CLS ensembles. This will not only enable a check of the lattice spacing and 
 (exponential) finite volume effects, but also elucidate the quark-mass 
 dependence by using ensembles along the $\mathrm{Tr} M = \mathrm{const}.$ 
 trajectory down to $m_{\pi} \lesssim 200\mathrm{MeV}$. While we have employed 
 the ground states in each irrep and a single excited state from the $(3,F_2)$ irrep here, at lighter pion masses, as $m_{\Delta}$ 
 moves further above the elastic threshold, more excited states will also be included to provide additional points. 

 This first elastic $\Delta(1232)\rightarrow N\pi$ calculation may be viewed 
 as a stepping stone in several respects. First, techniques for computing 
 correlation functions, extracting finite volume energies, and analysing determinant conditions may be extended to other resonant meson-baryon systems. 
 Such systems of interest may present additional complications like
 coupled scattering channels, as are present when studying the $\Lambda(1405)$ resonance. 
 Finally, building upon this calculation of $\Delta(1232)\rightarrow N\pi$ 
 will ultimately enable lattice QCD calculations of $\Delta(1232)$ transition 
 form-factors, for which the theoretical foundations can be found in 
 Ref.~\cite{Agadjanov:2014kha}. 
\begin{acknowledgments}
We acknowledge helpful discussions with Harvey Meyer and Daniel 
	Mohler. 
	This work was supported by the U.S. National Science 
	Foundation under Award No. PHY-1613449. 
	Some of our computations are performed using the CHROMA software suite~\cite{Edwards:2004sx}.  
	Computing facilities
	were provided by the Danish e-Infrastructure Cooperation (DeIC) National HPC
Centre at the University of Southern Denmark. We are grateful to our CLS 
	colleagues for sharing the gauge field configurations on
which this work is based. 
We acknowledge PRACE for awarding us access to resource FERMI based in Italy
at CINECA, Bologna and to resource SuperMUC based in Germany at LRZ, Munich.
Furthermore, this work was supported by a grant from the Swiss National Supercomputing
Centre (CSCS) under project ID s384. We are grateful for the support received by the
computer centers.
The authors gratefully acknowledge the Gauss Centre for Supercomputing (GCS)
for providing computing time through the John von Neumann Institute for Computing
	(NIC) on the GCS share of the supercomputer JUQUEEN at J\"{u}lich Supercomputing
Centre (JSC). GCS is the alliance of the three national supercomputing centres HLRS
	(Universit\"{a}t Stuttgart), JSC (Forschungszentrum J\"{u}lich), and LRZ (Bayerische Akademie
der Wissenschaften), funded by the German Federal Ministry of Education and 
	Research (BMBF) and the German State Ministries for Research of Baden-W\"{u}rttemberg (MWK), Bayern (StMWFK) and Nordrhein-Westfalen (MIWF).
\end{acknowledgments}

\bibliography{latticen}

\end{document}